\documentclass[a4paper]{article}
\usepackage{graphicx}
\usepackage[compress]{cite}
\usepackage{braket}
\usepackage{bm}
\usepackage{comment}
\usepackage{here}
\usepackage{a4wide}
\usepackage{authblk}
\usepackage{amsmath,amssymb}
\usepackage[usenames,dvipsnames]{color} 
\usepackage[normalem]{ulem} 
\usepackage{verbatim}

\begin{document}
\def\Journal#1#2#3#4{{#1} {\bf #2}, #3 (#4)}
\def\AHEP{Advances in High Energy Physics.} 	
\def\ARNPS{Annu. Rev. Nucl. Part. Sci.} 
\def\AandA{Astron. Astrophys.} 
\def\ANP{Ann. Phys.}
\def\APJ{Astrophys. J.}
\def\APJL{Astrophys. J. Lett.}
\def\APJS{Astrophys. J. Suppl}
\def\CMP{Commn. Math. Phys.}
\def\COMR{Comptes Rendues}
\def\CQG{Class. Quantum Grav.}
\def\CPC{Chin. Phys. C}
\def\EPJC{Eur. Phys. J. C}
\def\EPL{EPL}
\def\FP{Fortsch. Phys.}
\def\IJMPA{Int. J. Mod. Phys. A}
\def\IJMPE{Int. J. Mod. Phys. E}
\def\JCAP{J. Cosmol. Astropart. Phys.}
\def\JHEP{J. High Energy Phys.}
\def\JETPL{JETP. Lett.}
\def\JETPUSSR{JETP (USSR)}
\def\JPG{J. Phys. G} 
\def\JPCS{J. Phys. Conf. Ser.} 
\def\JPGNP{J. Phys. G: Nucl. Part. Phys.} 
\def\MNRAS{Mon. Not. R. Astron. Soc.}
\def\MPLA{Mod. Phys. Lett. A}
\def\NIMA{Nucl. Instrum. Meth. A.}
\def\NATU{Nature}
\def\NCA{Nuovo Cimento}
\def\NJP{New. J. Phys.}
\def\NPB{Nucl. Phys. B}
\def\NPBOLD{Nucl. Phys.}
\def\NPBSUPPL{Nucl. Phys. B. Proc. Suppl.}
\def\PDU{Phys. Dark. Univ.}
\def\PL{Phys. Lett.}
\def\PLB{{Phys. Lett.} B}
\def\PMCA{PMC Phys. A}
\def\PREP{Phys. Rep.}
\def\PPNP{Prog. Part. Nucl. Phys.}
\def\PLBOLD{Phys. Lett.}
\def\PAN{Phys. Atom. Nucl.}
\def\PRL{Phys. Rev. Lett.}
\def\PRD{Phys. Rev. D}
\def\PRC{Phys. Rev. C}
\def\PR{Phys. Rev.}
\def\PTP{Prog. Theor. Phys.}
\def\PTEP{Prog. Theor. Exp. Phys.}
\def\RMP{Rev. Mod. Phys.}
\def\RPP{Rep. Prog. Phys.}
\def\SJNP{Sov. J. Nucl. Phys.}
\def\SPJETP{Sov. Phys. JETP.}
\def\SCIENCE{Science}
\def\TNYAS{Trans. New York Acad. Sci.}
\def\ZETP{Zh. Eksp. Teor. Piz.}
\def\ZFPH{Z. fur Physik}
\def\ZPC{Z. Phys. C}
\title{Dirac mass matrix textures and the lightest right-handed neutrino mass scale in Type I seesaw leptogenesis}
\author[1]{Shuta Kosuge \footnote{5CSNM008@tokai.ac.jp}}
\author[2]{Teruyuki Kitabayashi \footnote{Corresponding author. teruyuki@tokai.ac.jp}}
\affil[1]{Graduate School of Science, Tokai University, 4-1-1 Kitakaname, Hiratsuka, Kanagawa 259-1292, Japan}
\affil[2]{Department of Physics, School of Science, Tokai University, 4-1-1 Kitakaname, Hiratsuka, Kanagawa 259-1292, Japan}
\date{}
\maketitle
\begin{abstract}
The type I seesaw mechanism is one of the leading proposed explanations for how neutrinos acquire their tiny masses. However, the mass scale of the undiscovered right-handed neutrinos required by this mechanism remains undetermined. Assuming vanilla leptogenesis in the two-flavor regime, we work backwards to find the required general textures of the Dirac mass matrix from which we determine the mass of the lightest right-handed neutrino to be around $10^9 {\rm GeV}$ to $10^{12} {\rm GeV}$.
\end{abstract}

\section{Introduction}
The origin of neutrino masses is one of the biggest unsolved problems in particle physics. One of the leading candidates for the generation of neutrino masses is the type I seesaw mechanism \cite{Minkowski1977PLB,Yanagida1979KEK,Gell-Mann1979,Glashow1980,Mohapatra1980}. The type I seesaw mechanism requires heavy right-handed neutrinos to exist, although these have not yet been discovered. Estimating the mass (energy) scale of right-handed neutrinos is crucial from a theoretical standpoint, as well as important for discovering them in experiments.

The heavy right-handed neutrino is also important as it may play a significant role in generating the baryon asymmetry of the universe via the proposed leptogenesis mechanism \cite{Fukugita1986PLB,Luty1992PRD,Covi1996PLB,Buchmuller1998PLB,E.Kh2003PLB,Guo2004PLB,Xing2021RPP}. According to this mechanism, depending on the mass scale of the lightest right-handed neutrino $M_1$, leptogenesis may proceed in one of three regimes \cite{Abada2006JCAP,Nardi2006JHEP}:
\begin{itemize} 
\item Unflavored regime, $M_1\gtrsim 10^{12} {\rm\: GeV}$: The Yukawa interactions of the three flavors of charged leptons cannot be distinguished from one another. Leptogenesis proceeds without distinguishing between the three flavors of charged leptons.
\item Two-flavor regime, $10^9 {\rm \:GeV} \lesssim M_1 \lesssim 10^{12} {\rm \:GeV}$: The Yukawa interactions of the tau can be distinguished from those of the other two charged leptons, the electron and muon.
\item Three-flavor regime, $M_1 \lesssim 10^9 {\rm\: GeV}$: All three flavors of charged leptons can be distinguished from one another.
\end{itemize}
Thinking about this in reverse, if we can determine in which regime leptogenesis occurs, we can determine the mass range of the lightest right-handed neutrino.

In this study, we show the relationship between the Dirac mass matrix and the right-handed neutrino mass in the context of a type I seesaw mechanism and leptogenesis scenario. We show that there are six general textures of the Dirac mass matrix which realize vanilla leptogenesis in the two-flavor regime. If the Dirac mass matrix takes on one of these particular forms, the mass of the lightest right-handed neutrino should be around $10^9 {\rm\: GeV}$ to $10^{12} {\rm \:GeV}$.

The choice of the two-flavor regime is physically significant for several reasons.  First, the mass scale of $10^9$ GeV to $10^{12}$ GeV is a critical transition zone where flavor effects begin to play a decisive role in the evolution of lepton asymmetry. From a theoretical viewpoint, the two-flavor regime represents the minimal extension beyond the unflavored approximation. In this regime, CP asymmetries and washout effects acquire nontrivial flavor structure, while the system remains analytically tractable. Furthermore, for hierarchical heavy neutrinos, $M_1 \simeq 10^9$ GeV is widely recognized as the lower bound for successful vanilla leptogenesis, known as the Davidson-Ibarra bound \cite{Davidson2002PLB}, making this specific range a benchmark for thermal leptogenesis models.

The remainder of this paper is organized as follows. Section 2 provides an overview of leptogenesis in the case of a type I seesaw mechanism. Section \ref{section_Dirac_mass_matrix} contains the main results of this paper. In section \ref{section_Dirac_general}, we show the general textures of the Dirac mass matrix which lead to an appropriate baryon asymmetry of the universe when leptogenesis is assumed to occur only in the two-flavor regime. Next, two specific textures of the Dirac mass matrix are discussed in sections \ref{section_specific1_mu_tau} and \ref{section_specific1_new}. The final section summarizes our work.

\section{Leptogenesis in Type I seesaw model \label{section_leptogenesis}}
In the Type I seesaw mechanism, the flavor neutrino mass matrix takes the following form
\begin{eqnarray}
M_\nu = -M_{\rm D} M_R^{-1} M_{\rm D}^T, 
\label{Eq:MnuSeesaw}
\end{eqnarray}
where
\begin{eqnarray}
M_{D}=\left( \begin{array}{ccc}
M_{e1} & M_{e2} & M_{e3} \\
M_{\mu1} & M_{\mu2} & M_{\mu 3} \\
M_{\tau1} & M_{\tau2} & M_{\tau 3}\\ 
\end{array} \right) 
=\left( \begin{array}{ccc}
r_{e1}e^{i\theta_{e1}} & r_{e2}e^{i\theta_{e2}} & r_{e3}e^{i\theta_{e3}} \\
r_{\mu1}e^{i\theta_{\mu 1}} & r_{\mu2}e^{i\theta_{\mu 2}} & r_{\mu 3}e^{i\theta_{\mu 3}}\\
r_{\tau1}e^{i\theta_{\tau 1}} & r_{\tau2}e^{i\theta_{\tau 2}} & r_{\tau 3}e^{i\theta_{\tau 3}}\\ 
\end{array} \right), 
\label{Eq:MD}
\end{eqnarray}
is the Dirac mass matrix with real parameters $r_{\alpha I}$ and $\theta_{\alpha I}$ ($I = 1,2,3$ and $\alpha = e, \mu, \tau$. $M_R = {\rm diag.}(M_1,M_2,M_3)$ denotes the mass matrix of the right-handed neutrinos. Eq (\ref{Eq:MD}) includes phases that can be eliminated by redefining the fields.

In the type I seesaw mechanism, lepton asymmetry arises from two reactions, $N_I\rightarrow L_\alpha+H$ and $N_I\rightarrow \bar{L}_\alpha+\bar{H}$, where $N_I$, $L$, and $H$ are the right-handed neutrinos, leptons, and Higgs fields, respectively. We consider only the case where the masses of right-handed neutrinos are highly hierarchical, namely $M_1 \ll M_2 \ll M_3$. In this case, the lepton asymmetry generated by the decay of $N_2$ and $N_3$ is erased as the universe evolves. Only the following asymmetry, which arises from the decay of $N_1$, survives
\begin{eqnarray}
\epsilon_{1\alpha} =\frac{1}{8\pi \left(M_{D}^{\dagger} M_{D} \right)_{11}v^{2}} \sum_{J=2}^3\left\{  \operatorname{Im}\left[X^{\alpha}_{1J}\right]F\left(\frac{M_J^2}{M_1^2}\right)+\operatorname{Im}\left[Y^{\alpha}_{1J}\right]G\left(\frac{M_J^2}{M_1^2}\right)\right\},  \label{4}
\end{eqnarray}
where
\begin{align}
X^{\alpha}_{1J}&=\left(M_{D}^{*} \right)_{\alpha 1} \left(M_{D} \right)_{\alpha J} \left(M_{D}^{\dagger}M_{D} \right)_{1J},  \label{X} \\
Y^{\alpha}_{1J}&=\left(M_{D}^{*} \right)_{\alpha 1} \left(M_{D} \right)_{\alpha J} \left(M_{D}^{\dagger}M_{D} \right)_{1J}^{*}, \label{Y} \\
F(x)&=\sqrt{x}\left(\frac{2-x}{1-x} +(1+x)\ln \left(\frac{x}{1+x}\right)\right), \\
G(x)&=\frac{1}{1-x},
\end{align}
and $v=174$ GeV denotes the Higgs vacuum expectation value. We note that $F(x) \simeq -3/(2\sqrt{x})$ and $G(x) \simeq -1/x$ for $x \gg 1$.

The lepton asymmetry that arises in the unflavored, two-flavor, and three-flavor regimes due to the $N_1$ decay are obtained as 
\begin{align}
Y_L^{(1)} & =  r(\epsilon_{1e}+\epsilon_{1\mu}+\epsilon_{1\tau}) \kappa (\tilde{m}_{1e}+\tilde{m}_{1\mu}+\tilde{m}_{1\tau}), \\
Y_L^{(2)} & = r\left[ (\epsilon_{1e}+\epsilon_{1\mu}) \kappa(\tilde{m}_{1e}+\tilde{m}_{1\mu}) + \epsilon_{1\tau} \kappa(\tilde{m}_{1\tau}) \right], \label{Eq:YL2}\\
Y_L^{(3)} & = r\left[ \epsilon_{1e} \kappa(\tilde{m}_{1e}) +\epsilon_{1\mu} \kappa(\tilde{m}_{1\mu}) + \epsilon_{1\tau} \kappa(\tilde{m}_{1\tau}) \right], 
\end{align}
respectively \cite{Shao2025PRD,Abada2006JCAP,Nardi2006JHEP}. Here, $r \simeq 3.9 \times 10^{-3}$ is the ratio of the particle density to the entropy density in the thermal equilibrium state of $N_1$,\cite{Xing2021RPP} and
\begin{eqnarray}
 \tilde{m}_{1\alpha}= \frac{(M_{D})_{\alpha 1} (M_{D})_{\alpha 1}^{*}}{M_1}. 
 \label{Eq:tildeM1alpha}
\end{eqnarray}
The dilution factor $\kappa$ represents the probability that the lepton asymmetry generated by the $N_1$ decay will survive. A calculation of the value of $\kappa$ relies on numerically solving a full set of Boltzmann equations, however, we use the following simpler empirical fit
formula for the efficiency factor \cite{Giudice2004NPB}
\begin{align}
\frac{1}{\kappa(\tilde{m}_1)} \simeq \frac{3.3 \times 10^{-3} ~{\rm eV}}{\tilde{m}_1} + \left( \frac{\tilde{m}_1}{5.5\times 10^{-4}~{\rm eV}}  \right)^{1.16}.
 \label{Eq:kappa}
\end{align}

In the leptogenesis scenario, the lepton number asymmetry arising from this CP asymmetry is converted into baryon asymmetry via the sphaleron process:
\begin{align}
Y_B = cY_L,
\label{Eq:YB=cYL}
\end{align}
where $c=28/79$ denotes conversion efficiency from the lepton asymmetry to the baryon asymmetry via the sphaleron processes.

\section{Dirac mass matrix textures for two-flavor regime\label{section_Dirac_mass_matrix}}
\subsection{General forms \label{section_Dirac_general}}
We now derive the Dirac mass matrix that leads to leptogenesis occurring in the two-flavor regime, i.e., a regime in which the mass range for right-handed neutrinos should be $10^9 {\rm GeV} \lesssim M_1 \lesssim 10^{12} {\rm GeV} $. Consider the case in $Y_L^{(1)} =0$, $Y_L^{(2)} \neq 0$, $Y_L^{(3)} = 0$, that is
\begin{align}
&\epsilon_{1e}+\epsilon_{1\mu}+\epsilon_{1\tau} =0, \label{Eq:Y1=0Y2neq0Y3=0_Y1=0}\\
& (\epsilon_{1e}+\epsilon_{1\mu}) \kappa(\tilde{m}_{1e}+\tilde{m}_{1\mu}) + \epsilon_{1\tau} \kappa(\tilde{m}_{1\tau}) \neq 0, \label{Eq:Y1=0Y2neq0Y3=0_Y2neq0}\\
& \epsilon_{1e} \kappa(\tilde{m}_{1e}) +\epsilon_{1\mu} \kappa(\tilde{m}_{1\mu}) + \epsilon_{1\tau} \kappa(\tilde{m}_{1\tau}) =0\label{Eq:Y1=0Y2neq0Y3=0_Y3=0}.
\end{align}

First, we assume that Eq.(\ref{Eq:Y1=0Y2neq0Y3=0_Y1=0}) holds. In this case,  Eq.(\ref{Eq:Y1=0Y2neq0Y3=0_Y3=0}) is satisfied if 
\begin{align}
\tilde{m}_{1e}=\tilde{m}_{1\mu}=\tilde{m}_{1\tau} \label{Eq:Y1=0Y2neq0Y3=0_tilde_m1e_tilde_1mu_tilde_1tau}.
\end{align}
This relationship is equivalent to
\begin{eqnarray}
r_{e1}^2 = r_{\mu 1}^2 = r_{\tau 1}^2,
\end{eqnarray}
from Eq.(\ref{Eq:tildeM1alpha}).

Eq.(\ref{Eq:Y1=0Y2neq0Y3=0_Y2neq0}) is satisfied if
\begin{align}
(\epsilon_{1e}+\epsilon_{1\mu}) \left[\kappa(2\tilde{m}_{1\tau}) -  \kappa(\tilde{m}_{1\tau})\right] \neq 0, 
\end{align}
and/or
\begin{align}
\epsilon_{1\tau} \left[\kappa(2\tilde{m}_{1\tau}) - \kappa(\tilde{m}_{1\tau})\right] \neq 0. 
\end{align}
These relationships are equivalent to
\begin{align}
\epsilon_{1e}+\epsilon_{1\mu}\neq 0,
\end{align}
or
\begin{align}
\epsilon_{1\tau}\neq 0,
\end{align}
with Eq.(\ref{Eq:Y1=0Y2neq0Y3=0_Y1=0}).

Therefore, the condition for leptogenesis to occur at $10^9 {\rm\: GeV} \lesssim M_1 \lesssim 10^{12} {\rm \:GeV} $ is
\begin{align}
& r_{\mu 1} = \rho r_{e1}, ~ r_{\tau 1} = \sigma r_{e1}, \label{Eq:2flavor_mtilder} \\
& \epsilon_{1e}+\epsilon_{1\mu}+\epsilon_{1\tau} =0,  \label{Eq:2flavor_epsilon} 
\end{align}
and
\begin{align}
\epsilon_{1e}+\epsilon_{1\mu}\neq 0 \quad {\rm or} \quad  \epsilon_{1\tau}\neq 0 \label{Eq:2flavor_epsilon-tau},
\end{align}
where $\rho = \pm 1$ and $\sigma = \pm 1$. The aim of this study will be achieved when we find the texture of the Dirac mass matrix that satisfies these conditions simultaneously.

First, we require the condition in Eq. (\ref{Eq:2flavor_mtilder}) to hold for the Dirac mass matrix in Eq.(\ref{Eq:MD}). Then, we have
\begin{eqnarray}
M_{D}=\left( \begin{array}{ccc}
r_{e1}e^{i\theta_{e1}} & r_{e2}e^{i\theta_{e2}} & r_{e3}e^{i\theta_{e3}} \\
\rho r_{e1}e^{i\theta_{\mu 1}} & r_{\mu2}e^{i\theta_{\mu 2}} & r_{\mu 3}e^{i\theta_{\mu 3}}\\
\sigma r_{e1}e^{i\theta_{\tau 1}} & r_{\tau2}e^{i\theta_{\tau 2}} & r_{\tau 3}e^{i\theta_{\tau 3}}\\ 
\end{array} \right).
\label{Eq:MD2}
\end{eqnarray}

Next, we also impose the condition in Eq.(\ref{Eq:2flavor_epsilon}) on the elements of the Dirac mass matrix in Eq.(\ref{Eq:MD2}). From Eq.(\ref{Eq:2flavor_epsilon}), we have
\begin{align}
   \varepsilon_{1e}+\varepsilon _{1\mu}+\varepsilon _{1\tau}
  &\propto 
  \sum_{J=2}^3\left\{  (\operatorname{Im}\left[X^{e}_{1J}\right]+\operatorname{Im}\left[X^{\mu}_{1J}\right]+\operatorname{Im}\left[X^{\tau}_{1J}\right])F\left(\frac{M_J^2}{M_1^2}\right) \right. \nonumber \\
  &\left. \quad  +(\operatorname{Im}\left[Y^{e}_{1J}\right]+\operatorname{Im}\left[Y^{\mu}_{1J}\right]+\operatorname{Im}\left[Y^{\tau}_{1J}\right])G\left(\frac{M_J^2}{M_1^2}\right) \right\} \nonumber \\
  &= 
  \sum_{J=2}^3\left\{  \operatorname{Im}\left[\left(M_{D}^{\dagger}M_{D} \right)_{1J}^{2}\right] F\left(\frac{M_J^2}{M_1^2}\right)\right.\notag \\
  &\left.\quad +\operatorname{Im}\left[\left(M_{D}^{\dagger}M_{D} \right)_{1J}\left(M_{D}^{\dagger}M_{D} \right)_{1J}^{*} \right]G\left(\frac{M_J^2}{M_1^2}\right)\right\} \nonumber \\
  & = 0.
\end{align}
Because  
\begin{align}
  \operatorname{Im}\left[Y^{e}_{1J}\right]+\operatorname{Im}\left[Y^{\mu}_{1J}\right]+\operatorname{Im}\left[Y^{\tau}_{1J}\right] =\operatorname{Im}\left[ \left|\left(M_{D}^{\dagger}M_{D} \right)_{1J}\right|^{2}   \right]=0,
\end{align}
the term proportional to $G$ becomes zero. The condition in Eq.\eqref{Eq:2flavor_epsilon} is satisfied when 
\begin{align}
\sum_{J=2}^3 \operatorname{Im}\left[\left(M_{D}^{\dagger}M_{D} \right)_{1J}^{2}\right] F\left(\frac{M_J^2}{M_1^2}\right) =2r_{e1}^{2}\sum_{J=2}^3({\bf s}_J\cdot {\bf r}_J)({\bf c}_J\cdot {\bf r}_J)F\left(\frac{M_J^2}{M_1^2}\right) = 0,
\label{Eq:Fpropto=0}
\end{align}
where
\begin{align}
  {\bf s}_J &=\left( \sin\phi_{eJ}, \sin\phi_{\mu J},  \sin\phi_{\tau J} \right), \notag \\
  {\bf c}_J &= \left( \cos\phi_{eJ}, \cos\phi_{\mu J},\cos\phi_{\tau J} \right), \notag \\
  {\bf r}_J  &= \left( r_{eJ}, \rho r_{\mu J}, \sigma r_{\tau J} \right),
\end{align}
and $\phi_{\alpha J}=\theta_{\alpha J}-\theta_{e1}$. There are three cases which satisfy Eq.(\ref{Eq:Fpropto=0}):

\begin{description}
\item[(i)] $r_{e1}=0$,
\item[(ii)] $({\bf s}_J\cdot {\bf r}_J)({\bf c}_J\cdot {\bf r}_J)=0$,
\item[(iii)] $\sum_{J=2}^3({\bf s}_J\cdot {\bf r}_J)({\bf c}_J\cdot {\bf r}_J)F\left(\frac{M_J^2}{M_1^2}\right)=0$.
\end{description}
In case (i), we obtain $\left(M_{D}^{\dagger} M_{D} \right)_{11} = 0$ and Eq\eqref{4} is not satisfied. Thus, the case (i) is excluded in our discussion. Moreover, we exclude the case (iii), because the condition in the case (iii) is dependent on the masses of the right-handed neutrinos via the function $F$. This conflicts with the aim of this study which is to determine the mass range of the lightest right-handed neutrino solely from the form of the Dirac mass matrix. Case (ii) is satisfied if ${\bf s}_J\cdot {\bf r}_J=0$ or ${\bf c}_J\cdot {\bf r}_J=0$. The first condition ${\bf s}_J\cdot {\bf r}_J=0$ is realized when
\begin{align}
  \left(\begin{matrix}
    r_{eJ} \\
    \rho r_{\mu J} \\
    \sigma r_{\tau J}
  \end{matrix}\right)=
  \alpha_J \left(\begin{matrix}
    \sin\phi_{\mu J} \\
    -\sin\phi_{eJ} \\
    0
  \end{matrix}\right)
  +
  \beta_J\left(\begin{matrix}
    \sin\phi_{\tau J} \\
    0 \\
    -\sin\phi_{eJ}
  \end{matrix}\right), \label{z2_1}
\end{align}
\begin{align}
  \left(\begin{matrix}
    r_{eJ} \\
    \rho r_{\mu J} \\
    \sigma r_{\tau J}
  \end{matrix}\right)=
  \alpha_J \left(\begin{matrix}
    \sin\phi_{\mu J} \\
    -\sin\phi_{eJ} \\
    0
  \end{matrix}\right)
  +
  \beta_J\left(\begin{matrix}
    0 \\
    -\sin\phi_{\tau J} \\
    \sin\phi_{\mu J}
  \end{matrix}\right), \label{z2_2}
\end{align}
or
\begin{align}
  \left(\begin{matrix}
    r_{eJ} \\
    \rho r_{\mu J} \\
    \sigma r_{\tau J}
  \end{matrix}\right)=
  \alpha_J \left(\begin{matrix}
    \sin\phi_{\tau J} \\
    0 \\
    -\sin\phi_{eJ}
      \end{matrix}\right)
  +
  \beta_J\left(\begin{matrix}
    0 \\
    \sin\phi_{\tau J} \\
    -\sin\phi_{\mu J}
  \end{matrix}\right) \label{z2_3},
\end{align}
where $\alpha_{J},\beta_{J} \in \mathbb{R} $. In these cases, the Dirac mass matrix may be

\begin{align}
  M_{D}^{\rm I}=
  \left(\begin{matrix}
    \vphantom{\displaystyle\frac{1}{\rho}(\alpha_2+\beta_2)}
    r_{e1}e^{i\theta_{e1}} & \begin{array}{c}(\alpha_2 \sin\theta_{\mu2 -\mu1} \\ \hspace{5em}+\beta_2 \sin\theta_{\tau2 -\tau1 })\\ \hspace{5em}\times e^{i\theta_{e2}}\\ \end{array} &\begin{array}{c}(\alpha_3 \sin\theta_{\mu3 -\mu1} \\ \hspace{5em}+\beta_3 \sin\theta_{\tau3 -\tau1 })\\ \hspace{5em}\times e^{i\theta_{e3}}\\ \end{array}\\[1ex]
    \vphantom{\displaystyle\frac{1}{\rho}(\alpha_2+\beta_2)} 
    \rho r_{e1}e^{i\theta_{\mu1}} & -\frac{\alpha_2}{\rho}\sin\theta_{e2-e1}e^{i\theta_{\mu2}} & -\frac{\alpha_3}{\rho}\sin\theta_{e3-e1}e^{i\theta_{\mu3}} \\[1ex]
    \vphantom{\displaystyle\frac{1}{\rho}(\alpha_2+\beta_2)}
    \sigma r_{e1}e^{i\theta_{\tau1}} & -\frac{\beta_2}{\sigma}\sin\theta_{e2-e1}e^{i\theta_{\tau2}} & -\frac{\beta_3}{\sigma}\sin\theta_{e3-e1}e^{i\theta_{\tau3}} 
    \end{matrix}\right),\label{MDs}
\end{align}

\begin{align}
  M_{D}^{\rm II}=\left(\begin{matrix}
    \vphantom{\displaystyle\frac{1}{\rho}(\alpha_2+\beta_2)}
    r_{e1}e^{i\theta_{e1}} & \alpha_2 \sin\theta_{\mu2 -\mu1}e^{i\theta_{e2}} & \alpha_3 \sin\theta_{\mu3 -\mu1}e^{i\theta_{e3}}\\[1ex]
    \vphantom{\displaystyle\frac{1}{\rho}(\alpha_2+\beta_2)}
    \rho r_{e1}e^{\mu1} &\begin{array}{c}-\frac{1}{\rho}(\alpha_2\sin\theta_{e2-e1} \\ \hspace{5em}+\beta_2\sin\theta_{\tau2-\tau1}) \\\hspace{5em} \times e^{i\theta_{\mu2}}\end{array} & \begin{array}{c}-\frac{1}{\rho}(\alpha_3\sin\theta_{e3-e1} \\ \hspace{5em}+\beta_3\sin\theta_{\tau3-\tau1}) \\\hspace{5em} \times e^{i\theta_{\mu3}}\end{array} \\[1ex]
     \vphantom{\displaystyle\frac{1}{\rho}(\alpha_2+\beta_2)}
     \sigma r_{e1}e^{i\theta_{\tau1}} & \frac{\beta_2}{\sigma} \sin\theta_{\mu2 -\mu1}e^{i\theta_{\tau2}} & \frac{\beta_3}{\sigma}\sin\theta_{\mu3 -\mu1}e^{i\theta_{\tau3}}\\
  \end{matrix} \right), \label{MDss}
\end{align}
or
\begin{align}
  M_{D}^{\rm III}=\left(\begin{matrix} \vphantom{\displaystyle\frac{1}{\rho}(\alpha_2+\beta_2)}
    r_{e1}e^{i\theta_{e1}} & \alpha_2 \sin\theta_{\tau2 -\tau1}e^{i\theta_{e2}} & \alpha_3 \sin\theta_{\tau3 -\tau1}e^{i\theta_{e3}}\\[1ex]
    \vphantom{\displaystyle\frac{1}{\rho}(\alpha_2+\beta_2)}
    \rho r_{e1}e^{i\theta_{\mu1}} & \frac{\beta_2}{\rho}\sin\theta_{\tau2-\tau1}e^{i\theta_{\mu2}} & \frac{\beta_3}{\rho}\sin\theta_{\tau3-\tau1}e^{i\theta_{\mu3}} \\[1ex]
    \vphantom{\displaystyle\frac{1}{\rho}(\alpha_2+\beta_2)}
    \sigma r_{e1}e^{i\theta_{\tau1}} & \begin{array}{c}-\frac{1}{\sigma}(\alpha_2 \sin\theta_{e2 -e1} \\ \hspace{5em}+\beta_2 \sin\theta_{\mu2 -\mu1}) \\ \hspace{5em} \times  e^{i\theta_{\tau2}}\end{array} & \begin{array}{c}-\frac{1}{\sigma}(\alpha_3 \sin\theta_{e3 -e1} \\ \hspace{5em}+\beta_3 \sin\theta_{\mu3 -\mu1}) \\ \hspace{5em} \times  e^{i\theta_{\tau3}}\end{array} 
  \end{matrix}\right), \label{MDsss}
\end{align}
where $\sin\theta_{\alpha i - \alpha j} = \sin(\theta_{\alpha i}-\theta_{\alpha j})$, e.g., $\sin\theta_{\mu 2 - \mu 1} = \sin(\theta_{\mu 2}-\theta_{\mu 2})$. Similarly, the condition of ${\bf c}_J\cdot {\bf r}_J=0$ is satisfied with Eqs.(\ref{z2_1}), (\ref{z2_2}) and (\ref{z2_3}) if we shift the phases by $\pi/2$. In these cases, the Dirac mass matrix may be

\begin{align}
  M_{D}^{\rm IV}=
  \left(\begin{matrix}
    \vphantom{\displaystyle\frac{1}{\rho}(\alpha_2+\beta_2)}
    r_{e1}e^{i\theta_{e1}} & \begin{array}{c}(\alpha_2 \cos\theta_{\mu2-\mu1}\\ \hspace{5em}+\beta_2\cos\theta_{\tau2-\tau1})\\ \hspace{5em} \times e^{i\theta_{e2}}\end{array} &\begin{array}{c}(\alpha_3 \cos\theta_{\mu3-\mu1}\\ \hspace{5em}+\beta_3\cos\theta_{\tau3-\tau1})\\ \hspace{5em} \times e^{i\theta_{e3}}\end{array} \\[1ex]
    \vphantom{\displaystyle\frac{1}{\rho}(\alpha_2+\beta_2)}
    \rho r_{e1}e^{i\theta_{\mu1}} & -\frac{\alpha_2}{\rho}\cos\theta_{e2-e1}e^{i\theta_{\mu2}} & -\frac{\alpha_3}{\rho}\cos\theta_{e3-e1}e^{i\theta_{\mu3}} \\[1ex]
    \vphantom{\displaystyle\frac{1}{\rho}(\alpha_2+\beta_2)}
    \sigma r_{e1}e^{i\theta_{\tau1}} & -\frac{\beta_2}{\sigma}\cos\theta_{e2-e1}e^{i\theta_{\tau2}} & -\frac{\beta_3}{\sigma}\cos\theta_{e3-e1}e^{i\theta_{\tau3}}
    \end{matrix}\right), \label{MDc}
\end{align}

\begin{align}
  M_{D}^{\rm V}=\left(\begin{matrix}\vphantom{\displaystyle\frac{1}{\rho}(\alpha_2+\beta_2)}
    r_{e1}e^{i\theta_{e1}} & \alpha_2 \cos\theta_{\mu2 -\mu1}e^{i\theta_{e2}} & \alpha_3 \cos\theta_{\mu3 -\mu1}e^{i\theta_{e3}}\\[1ex]
    \vphantom{\displaystyle\frac{1}{\rho}(\alpha_2+\beta_2)}
    \rho r_{e1}e^{i\theta_{\mu1}} & \begin{array}{c}-\frac{1}{\rho}(\alpha_2 \cos\theta_{e2 -e1}\\ \hspace{5em}+\beta_2 \cos\theta_{\tau2 -\tau1}) \\ \hspace{5em}\times e^{i\theta_{\mu2}}\end{array} & \begin{array}{c}-\frac{1}{\rho}(\alpha_3 \cos\theta_{e3 -e1}\\ \hspace{5em}+\beta_3 \cos\theta_{\tau3 -\tau1}) \\ \hspace{5em}\times e^{i\theta_{\mu3}}\end{array}\\[1ex] 
    \vphantom{\displaystyle\frac{1}{\rho}(\alpha_2+\beta_2)}
    \sigma r_{e1}e^{i\theta_{\tau1}} & \frac{\beta_2}{\sigma}\cos\theta_{\mu2-\mu1}e^{i\theta_{\tau2}} & \frac{\beta_3}{\sigma}\cos\theta_{\mu3-\mu1}e^{i\theta_{\tau3}}
  \end{matrix}\right), \label{MDcc}
\end{align}
or
\begin{align}
  M_{D}^{\rm VI}=\left(\begin{matrix}\vphantom{\displaystyle\frac{1}{\rho}(\alpha_2+\beta_2)}
    r_{e1}e^{i\theta_{e1}} & \alpha_2 \cos\theta_{\tau2 -\tau1}e^{i\theta_{e2}} & \alpha_3 \cos\theta_{\tau3 -\tau1}e^{i\theta_{e3}}\\[1ex]
    \vphantom{\displaystyle\frac{1}{\rho}(\alpha_2+\beta_2)}
    \rho r_{e1}e^{i\theta_{\mu1}} & \frac{\beta_2}{\rho}\cos\theta_{\tau2-\tau1}e^{i\theta_{\mu2}} & \frac{\beta_3}{\rho}\cos\theta_{\tau3-\tau1}e^{i\theta_{\mu3}} \\[1ex]
    \vphantom{\displaystyle\frac{1}{\rho}(\alpha_2+\beta_2)}
    \sigma r_{e1}e^{i\theta_{\tau1}} & \begin{array}{c}-\frac{1}{\sigma}(\alpha_2 \cos\theta_{e2 -e1}\\ \hspace{5em}+\beta_2 \cos\theta_{\mu2 -\mu1})\\ \hspace{5em} \times e^{i\theta_{\tau2}} \end{array} & \begin{array}{c}-\frac{1}{\sigma}(\alpha_3 \cos\theta_{e3-e1}\\ \hspace{5em}+\beta_3 \cos\theta_{\mu3 -\mu1})\\ \hspace{5em} \times e^{i\theta_{\tau3}} \end{array}
  \end{matrix}\right), \label{MDccc}
\end{align}
where $\cos\theta_{\alpha i - \beta j} = \cos(\theta_{\alpha i}-\theta_{\beta j})$. The resulting $\epsilon_{1\alpha}$ from $M_{D}^{\rm I}$, $M_{D}^{\rm II}$, $\cdots$, $M_{D}^{\rm VI}$ are obtained as follows.

\begin{description}
\item[ $M_{D}^{\rm I}$:] 
\begin{align}
\epsilon_{1e}  &=\sum_{J=2}^{3} \epsilon_{ 1e}^{(J)}=\sum_{J=2}^3 \frac{1}{24\pi v^{2}}\sin\theta_{eJ-e1}(\alpha_J \sin\theta_{\mu J-\mu 1}+\beta_J \sin\theta_{\tau J-\tau 1}) \notag \\
& \hspace{6em} \times (\alpha_J\sin\theta_{\mu J-\mu 1-eJ+e1}+\beta_J\sin\theta_{\tau J-\tau1-eJ+e1}) \notag \\
& \hspace{6em} \times \left\{F\left(\frac{M_J^2}{M_1^2}\right)+G\left(\frac{M_J^2}{M_1^2}\right)\right\}   \nonumber \\ 
\epsilon_{1\mu}  &=-\sum_{J=2}^3 \frac{\alpha_J \sin\theta_{\mu J-\mu 1}}{\alpha_J \sin\theta_{\mu J-\mu 1}+\beta_J \sin\theta_{\tau J-\tau 1}} \epsilon_{1e}^{(J)} \nonumber \\ 
\epsilon_{1\tau} & =-\sum_{J=2}^3\frac{\beta_J \sin\theta_{\tau J-\tau 1}}{\alpha_J \sin\theta_{\mu J-\mu 1}+\beta_J \sin\theta_{\tau J-\tau 1}} \epsilon_{1e}^{(J)} 
\end{align}

\item[ $M_{D}^{\rm II}$:] 
\begin{align}
\epsilon_{1e} & =-\sum_{J=2}^3\frac{\alpha_J \sin\theta_{e J-e1}}{\alpha_J \sin\theta_{e J-e1}+\beta_J \sin\theta_{\tau J-\tau1}}\epsilon_{1\mu}^{(J)}   \nonumber \\ 
\epsilon_{1\mu} & =\sum_{J=2}^{3} \epsilon_{ 1\mu}^{(J)}=\sum_{J=2}^3\frac{1}{24\pi v^{2}}\sin\theta_{\mu J-\mu1}(\alpha_J \sin\theta_{e J-e1}+\beta_J \sin\theta_{\tau J-\tau1})   \notag \\
& \hspace{6em} \times (\alpha_J\sin\theta_{e J-e 1-\mu J+\mu1}+\beta_J\sin\theta_{\tau J-\tau1-\mu J+\mu 1}) \notag \\
& \hspace{6em} \times \left\{F\left(\frac{M_J^2}{M_1^2}\right)+G\left(\frac{M_J^2}{M_1^2}\right)\right\} \nonumber \\
\epsilon_{1\tau} & =  -\sum_{J=2}^3\frac{\beta_J \sin\theta_{\tau J-\tau1}}{\alpha_J \sin\theta_{e J-e1}+\beta_J \sin\theta_{\tau J-\tau1}}\epsilon_{1\mu}^{(J)}
\end{align}

\item[ $M_{D}^{\rm III}$:] 
\begin{align}
\epsilon_{1e} & =-\sum_{J=2}^3\frac{\alpha_J \sin\theta_{e J-e1}}{\alpha_J \sin\theta_{e J-e1}+\beta_J \sin\theta_{\mu J-\mu1}}\epsilon_{1\tau}^{(J)}  \nonumber \\ 
\epsilon_{1\mu} & =-\sum_{J=2}^3\frac{\beta_J \sin\theta_{\mu J-\mu1}}{\alpha_J \sin\theta_{e J-e1}+\beta_J \sin\theta_{\mu J-\mu1}}\epsilon_{1\tau}^{(J)}  \nonumber \\
\epsilon_{1\tau} & =\sum_{J=2}^{3} \epsilon_{ 1\tau}^{(J)}= \sum_{J=2}^3\frac{1}{24\pi v^{2}}\sin\theta_{\tau J-\tau1}(\alpha_J \sin\theta_{e J-e1}+\beta_J \sin\theta_{\mu J-\mu1})   \notag \\
& \hspace{6em}\times (\alpha_J\sin\theta_{e J-e 1-\tau J+\tau1}+\beta_J\sin\theta_{\mu J-\mu1-\tau J+\tau 1}) \notag \\
&\hspace{6em}\times \left\{F\left(\frac{M_J^2}{M_1^2}\right)+G\left(\frac{M_J^2}{M_1^2}\right)\right\}
\end{align}

\item[ $M_{D}^{\rm IV}$:] 
\begin{align}
\epsilon_{1e} & =\sum_{J=2}^{3} \epsilon_{ 1e}^{(J)}=\sum_{J=2}^3 \frac{-1}{24\pi v^{2}}\cos\theta_{eJ-e1}(\alpha_J \cos\theta_{\mu J-\mu 1}+\beta_J \cos\theta_{\tau J-\tau 1}) \notag \\
&\hspace{6em}\times (\alpha_J\sin\theta_{\mu J-\mu 1-eJ+e1}+\beta_J\sin\theta_{\tau J-\tau1-eJ+e1}) \notag \\
& \hspace{6em} \times \left\{F\left(\frac{M_J^2}{M_1^2}\right)-G\left(\frac{M_J^2}{M_1^2}\right)\right\}  \nonumber \\ 
\epsilon_{1\mu} & =\sum_{J=2}^3\frac{\alpha_J \cos\theta_{\mu J-\mu 1}}{\alpha_J \cos\theta_{\mu J-\mu 1}+\beta_J \cos\theta_{\tau J-\tau 1}} \epsilon_{1e}^{(J)}  \nonumber \\
\epsilon_{1\tau} & = \sum_{J=2}^3\frac{\beta_J \cos\theta_{\tau J-\tau 1}}{\alpha_J \cos\theta_{\mu J-\mu 1}+\beta_J \cos\theta_{\tau J-\tau 1}} \epsilon_{1e}^{(J)}
\end{align}

\item[ $M_{D}^{\rm V}$:] 
\begin{align}
\epsilon_{1e} & =\sum_{J=2}^3\frac{\alpha_J \cos\theta_{e J-e1}}{\alpha_J \cos\theta_{e J-e1}+\beta_J \cos\theta_{\tau J-\tau1}}\epsilon_{1\mu}^{(J)}  \nonumber \\ 
\epsilon_{1\mu} & =\sum_{J=2}^{3} \epsilon_{ 1\mu}^{(J)}=\sum_{J=2}^3\frac{-1}{24\pi v^{2}}\cos\theta_{\mu J-\mu1}(\alpha_J \cos\theta_{e J-e1}+\beta_J \cos\theta_{\tau J-\tau1})   \notag \\
& \hspace{6em}\times (\alpha_J\sin\theta_{e J-e 1-\mu J+\mu1}+\beta_J\sin\theta_{\tau J-\tau1-\mu J+\mu 1})\notag \\
&\hspace{6em}\times \left\{F\left(\frac{M_J^2}{M_1^2}\right)-G\left(\frac{M_J^2}{M_1^2}\right)\right\}  \nonumber \\
\epsilon_{1\tau} & =\sum_{J=2}^3 \frac{\beta_J \cos\theta_{\tau J-\tau1}}{\alpha_J \cos\theta_{e J-e1}+\beta_J \cos\theta_{\tau J-\tau1}}\epsilon_{1\mu}^{(J)}
\end{align}

\item[ $M_{D}^{\rm VI}$:] 
\begin{align}
\epsilon_{1e} & =\sum_{J=2}^3\frac{\alpha_J \cos\theta_{e J-e1}}{\alpha_J \cos\theta_{e J-e1}+\beta_J \cos\theta_{\mu J-\mu1}}\epsilon_{1\tau}^{(J)} \nonumber \\ 
\epsilon_{1\mu} & =\sum_{J=2}^3\frac{\beta_J \cos\theta_{\mu J-\mu1}}{\alpha_J \cos\theta_{e J-e1}+\beta_J \cos\theta_{\mu J-\mu1}}\epsilon_{1\tau}^{(J)}   \nonumber \\
\epsilon_{1\tau} & =\sum_{J=2}^{3} \epsilon_{ 1\tau}^{(J)}= \sum_{J=2}^3\frac{-1}{24\pi v^{2}}\cos\theta_{\tau J-\tau1}(\alpha_J \cos\theta_{e J-e1}+\beta_J \cos\theta_{\mu J-\mu1})   \notag \\
& \hspace{6em}\times (\alpha_J\sin\theta_{e J-e 1-\tau J+\tau1}+\beta_J\sin\theta_{\mu J-\mu1-\tau J+\tau 1}) \notag \\
&\hspace{6em} \times \left\{F\left(\frac{M_J^2}{M_1^2}\right)-G\left(\frac{M_J^2}{M_1^2}\right)\right\} 
\end{align}
\end{description}
Here $\sin\theta_{\alpha J-\alpha 1 - \beta J+\beta 1} = \sin(\theta_{\alpha J}-\theta_{\alpha 1}-\theta_{\beta J}+\theta_{\beta 1})$ and $\cos\theta_{\alpha J-\alpha 1 - \beta J+\beta 1} = \cos(\theta_{\alpha J}-\theta_{\alpha 1}-\theta_{\beta J}+\theta_{\beta 1})$. The resulting $\tilde{m}_{1\alpha}$ from $M_{D}^{\rm I}$, $M_{D}^{\rm II}$, $\cdots$, $M_{D}^{\rm VI}$ all satisfy 
\begin{align}
\tilde{m}_{1e}  = \tilde{m}_{1\mu}  =  \tilde{m}_{1\tau}  =  \frac{r_{e1}^{2}}{M_{1}}.
\end{align}

Finally, we impose the condition in Eq.(\ref{Eq:2flavor_epsilon-tau}) on the elements of the Dirac mass matrix. Since 
\begin{eqnarray}
\frac{G(x)}{F(x)} \simeq \frac{2}{3}\frac{M_1}{M_J} \ll 1,
\end{eqnarray}
for $x=M_J^2/M_1^2 \gg 1$, we omit the term proportional to $G$ in Eq.(\ref{4}). In addition, because of 
\begin{eqnarray}
\frac{F\left(\frac{M_3^2}{M_1^2}\right)}{ F\left(\frac{M_2^2}{M_1^2}\right)} =  \frac{M_2}{M_3} \ll 1,
\end{eqnarray}
we omit $F\left(\frac{M_3^2}{M_1^2}\right)$. Then $\epsilon_{1e}+\epsilon_{1\mu}\neq 0$ and $\epsilon_{1\tau} \neq 0$ becomes
\begin{align}
\operatorname{Im}\left[X^{e}_{12}\right]+\operatorname{Im}\left[X^{\mu}_{12}\right]\neq 0
 \label{Eq:ImXeXmu_neq_0}
\end{align}
and
\begin{align}
\operatorname{Im}\left[X^{\tau}_{12}\right]\neq 0, \label{Eq:ImXtau_neq_0}
\end{align}
respectively. As we mentioned, it is enough that only \eqref{Eq:ImXeXmu_neq_0} or \eqref{Eq:ImXtau_neq_0} is satisfied. We choose the condition in Eq.(\ref{Eq:ImXtau_neq_0}). For $M_{D}^{\rm I}$, $M_{D}^{\rm II}$, $\cdots$, $M_{D}^{\rm VI}$, Eq.(\ref{Eq:ImXtau_neq_0}) is satisfied when 
\begin{align}
  r_{e1}\neq 0, \quad \beta_2\neq 0,
\end{align}
and the following relations ($n=0,1,2,\cdots$) hold.
\begin{description}
\item[ $M_{D}^{\rm I}:$] 
\begin{align}
\theta_{e 2-e 1} \neq n\pi, \quad \theta_{\tau 2-\tau 1} \neq n\pi,\quad
 \frac{\alpha_2}{\beta_2}    \neq -\frac{\sin\theta_{\tau2-\tau1 -e 2 + e 1} }{\sin\theta_{\mu2-\mu1 -e 2 + e 1} }.
\label{MD_I_conditions1}
\end{align}
\item[ $M_{D}^{\rm II}:$] 
\begin{align}
\theta_{\mu 2-\mu 1} \neq n\pi, \quad \theta_{\tau 2-\tau 1} \neq n\pi,\quad
 \frac{\alpha_2}{\beta_2}    \neq -\frac{\sin\theta_{\tau2-\tau1 -\mu 2 + \mu 1} }{\sin\theta_{e2-e1 -\mu 2 + \mu 1} }.
\label{MD_II_conditions1}
\end{align}

\item[ $M_{D}^{\rm III}:$] 
\begin{align}
\theta_{\tau 2-\tau 1} \neq n\pi, \quad  \frac{\alpha_2}{\beta_2} \neq -\frac{\sin\theta_{\mu2-\mu1}}{\sin\theta_{e2-e1}},\quad
 \frac{\alpha_2}{\beta_2}    \neq -\frac{\sin\theta_{\mu2-\mu1 -\tau 2 + \tau 1} }{\sin\theta_{e2-e1 -\tau 2 + \tau 1} }.
\label{MD_III_conditions1}
\end{align}

\item[ $M_{D}^{\rm IV}:$] 
\begin{align}
\theta_{e 2-e 1} \neq \left(\frac{1}{2}+n\right)\pi, \quad \theta_{\tau 2-\tau 1} \neq \left(\frac{1}{2}+n\right)\pi,\quad
 \frac{\alpha_2}{\beta_2}    \neq -\frac{\sin\theta_{\tau2-\tau1 -e 2 + e 1} }{\sin\theta_{\mu2-\mu1 -e 2 + e 1} }.
\label{MD_IV_conditions1}
\end{align}

\item[ $M_{D}^{\rm V}:$] 
\begin{align}
\theta_{\mu 2-\mu 1} \neq \left(\frac{1}{2}+n\right)\pi, \quad \theta_{\tau 2-\tau 1} \neq \left(\frac{1}{2}+n\right)\pi,\quad
 \frac{\alpha_2}{\beta_2}    \neq -\frac{\sin\theta_{\tau2-\tau1 -\mu 2 + \mu 1} }{\sin\theta_{e2-e1 -\mu 2 + \mu 1} }.
\label{MD_V_conditions1}
\end{align}

\item[ $M_{D}^{\rm VI}:$] 
\begin{align}
\theta_{\tau 2-\tau 1} \neq \left(\frac{1}{2}+n\right)\pi , \quad  \frac{\alpha_2}{\beta_2} \neq -\frac{\cos\theta_{\mu2-\mu1}}{\cos\theta_{e2-e1}},\quad
 \frac{\alpha_2}{\beta_2}    \neq -\frac{\sin\theta_{\mu2-\mu1 -\tau 2 + \tau 1} }{\sin\theta_{e2-e1 -\tau 2 + \tau 1} }.
\label{MD_VI_conditions1}
\end{align}
\end{description}

The main aim of this study has been achieved. For example, it can be seen that leptogenesis occurs only in the two-flavor regime when the Dirac mass matrix is given by $M_{D}^{\rm I}$ in Eq.\eqref{MDs} and the condition in Eq.\eqref{MD_I_conditions1} holds. In this case, the mass range of the lightest right-handed neutrino must be $10^{9} \text{ GeV} \lesssim M_1 \lesssim 10^{12} \text{ GeV}$. Similar conclusions are obtained for $M_{D}^{\rm II}$, $M_{D}^{\rm III}$, $\cdots$, $M_{D}^{\rm VI}$. 

Up to now, we have included the global phase that can be eliminated by redefining the fields in the Dirac mass matrix. To derive the most general texture of mass matrix, we have also not considered the number of parameters allowed in the elementary particle model underlying the Dirac mass matrix. However, in practice, the non-physical phases within the Dirac mass matrix can be removed. The method of removing phases is arbitrary, and various phase redefinitions are adopted in different models. Furthermore, the number of allowed free parameters also varies depending on the particle content of the particular type I seesaw model being employed.

In the remainder of this section, we present two specific cases of the Dirac mass matrix which lead to the lightest right-handed neutrino mass range being $10^{9} \text{ GeV} \lesssim M_1 \lesssim10^{12} \text{ GeV}$.

\subsection{Specific case 1: Comparison with a previous study\label{section_specific1_mu_tau}}
It is known that, given the following form of the Dirac mass matrix, that only two-flavor leptogenesis is allowed
\cite{Shao2025PRD}
\begin{eqnarray}
\left( \begin{array}{ccc}
a & b & c \\
a' & b' & c' \\
a'^* & b'^* & c'^* \\
\end{array} \right) 
\left( \begin{array}{ccc}
\sqrt{\eta_1}& 0 & 0 \\
0 & \sqrt{\eta_2} & 0 \\
0 & 0 & \sqrt{\eta_3} \\
\end{array} \right),
\label{Eq:MD_mu_tau}
\end{eqnarray}
where $a$, $b$ and $c$ ($a'$, $b'$, and $c'$) are real (complex) elements and $\eta_I=\pm 1$. 

The Dirac mass matrix in Eq.\eqref{Eq:MD_mu_tau} is a specific form of $M_{D}^{\rm II}$ in Eq.\eqref{MDss}. Indeed, all components in the first row in $M_{D}^{\rm II}$ can be real parameters if we set $\theta_{e1} = \theta_{e2} = \theta_{e3} = \theta$ and the fields are redefined such as to absorb the common phase $\theta$. As a result of this redefinition, the phase of the remaining elements shifts. We then redefine this shifted phase. For example, we rewrite phase of the (2,1) component from $\theta_{\mu 1}-\theta$ as $\theta_{\mu 1}$. Then we let $\rho=\sigma$, $-\theta_{\tau 1} =  \theta_{\mu 1} =  \theta_1$, $-\theta_{\tau 2} =  \theta_{\mu 2} =  \theta_2$, $-\theta_{\tau 3} =  \theta_{\mu 3} =  \theta_3$ and $r_{e1}=A$ with$\theta_2 \neq \theta_1$. The Dirac mass matrix then becomes 
\begin{align}
  M_{D}^{\rm II}=
  \left(\begin{matrix}
    A & \alpha_2 \sin(\theta_2-\theta_1) & \alpha_3 \sin(\theta_3-\theta_1) \\
    \rho A e^{i\theta_1} & \frac{1}{\rho}\beta_2\sin(\theta_2-\theta_1) e^{i\theta_2} & \frac{1}{\rho}\beta_3\sin(\theta_3-\theta_1) e^{i\theta_3} \\
    \rho A e^{-i\theta_1} & \frac{1}{\rho}\beta_2\sin(\theta_2-\theta_1) e^{-i\theta_2} & \frac{1}{\rho}\beta_3\sin(\theta_3-\theta_1) e^{-i\theta_3} 
    \end{matrix}\right) 
=   
 \left( \begin{array}{ccc}
a & b & c \\
a' & b' & c' \\
a'^* & b'^* & c'^* \\
\end{array} \right), 
\label{Eq:MII_mu_tau}
\end{align}
where we make several substitutions such as $A=a$ and $ Ae^{i\theta_1} = a'$ for readability. Eq.\eqref{Eq:MII_mu_tau} is same of Eq.\eqref{Eq:MD_mu_tau} for $\eta_I=+1$.

For the sake of completeness, let us explicitly confirm that the Dirac mass matrix shown in Eq.(\ref{Eq:MII_mu_tau}) permits only the two-flavor regime (for which the mass of the lightest right-handed neutrino should be $10^{9} \text{ GeV} \leq M_1 \leq 10^{12} \text{ GeV}$). From Eq.(\ref{Eq:MII_mu_tau}), we obtain

\begin{align}
\epsilon_{1e} & =0, \nonumber \\ 
\epsilon_{1\mu} & =\frac{1}{24\pi v^{2}}\sum_{J=2}^3  \beta_J  \sin^{2}(\theta_J-\theta_1)[\alpha_J\sin(\theta_J-\theta_1)+ \beta_J\sin2(\theta_J-\theta_1)] \notag \\
& \quad \times \left\{F\left(\frac{M_J^2}{M_1^2}\right)+G\left(\frac{M_J^2}{M_1^2}\right)\right\}, \notag \\
\epsilon_{1\tau} & = -\epsilon_{1\mu}, \label{epsilon123}
\end{align}
and $\tilde{m}_{1e}  = \tilde{m}_{1\mu}  = \tilde{m}_{1\tau}  =A^2/M_1$. Therefore, Eqs. (\ref{Eq:Y1=0Y2neq0Y3=0_Y1=0}),  (\ref{Eq:Y1=0Y2neq0Y3=0_Y2neq0}) and (\ref{Eq:Y1=0Y2neq0Y3=0_Y3=0}) hold and leptogenesis only occurs in two-flavor regime. In this case, the lightest right-handed neutrino mass range should be $10^{9} \text{ GeV} \lesssim M_1 \lesssim 10^{12} \text{ GeV}$.\newline

\subsection{Specific case 2: a new texture\label{section_specific1_new}}
Next, we show a new specific texture of the Dirac mass matrix for $10^{9} \text{ GeV} \leq M_1 \leq 10^{12} \text{ GeV}$. In $M_{D}^{\rm IV}$ let $\theta_{e1}=\theta_{e2}=\theta_{e3}=\theta$. Then, similar to Eq.(\ref{Eq:MII_mu_tau}), the common phase is absorbed into the field. Next, we set $\theta_{\tau 1} = \theta_{\mu 1} = \theta_1$, $\theta_{\tau 2} = \theta_{\mu 2} = \theta_2$, and $\theta_{\tau 3} = \theta_{\mu 3} = \theta_3$. In this case, the Dirac mass matrix which is obtained as 
\begin{align}
  M_{D}^{\rm IV}=
  \left(\begin{matrix}
    A & 2\alpha_2 \cos(\theta_2-\theta_1) & 2\alpha_3 \cos(\theta_3-\theta_1) \\
    A e^{i\theta_1} & -\alpha_2 e^{i\theta_2} & -\alpha_3 e^{i\theta_3} \\
    A e^{i\theta_1} & -\alpha_2 e^{i\theta_2} & -\alpha_3 e^{i\theta_3}
    \end{matrix}\right) 
\label{Eq:MII_new}.
\end{align}
From Eq.(\ref{Eq:MII_new}) we derive
\begin{align}
\epsilon_{1e} & = -\frac{1}{12 \pi v^{2}} \sum_{J=2}^3 \alpha_{J}^{2} \sin2(\theta_{J}-\theta_{1})   \left\{F\left(\frac{M_J^2}{M_1^2}\right)-G\left(\frac{M_J^2}{M_1^2}\right)\right\},   \nonumber \\ 
\epsilon_{1\mu} & = \epsilon_{1\tau}=-\frac{\epsilon_{1e}}{2},
\label{Eq:epsilon_in_new_model}
\end{align}
and $\tilde{m}_{1e}  = \tilde{m}_{1\mu}  = \tilde{m}_{1\tau}  =A^2/M_1$, and Eqs.(\ref{Eq:Y1=0Y2neq0Y3=0_Y1=0}),  (\ref{Eq:Y1=0Y2neq0Y3=0_Y2neq0}) and (\ref{Eq:Y1=0Y2neq0Y3=0_Y3=0}) are satisfied. Thus, leptogenesis can work only in the two-flavor regime and the lightest right-handed neutrino mass range should be $10^{9} \text{ GeV} \lesssim M_1 \lesssim 10^{12} \text{ GeV}$.

\subsection{Possibility of successful leptogenesis}
Although our primary focus is on the algebraic derivation of the Dirac mass matrix textures that isolate the two-flavor regime,  $10^9~\mathrm{GeV} \lesssim M_1 \lesssim 10^{12}~\mathrm{GeV}$,  it is important to examine whether the textures derived in this study can reproduce the observed baryon asymmetry of the Universe \cite{Planck2020AA}, $Y_B \simeq 8.7 \times 10^{-11}$.

For example, we estimate the baryon asymmetry for the new texture introduced in Section~\ref{section_specific1_new}. From Eqs.~(\ref{Eq:YL2}), (\ref{Eq:kappa}), (\ref{Eq:YB=cYL}), and (\ref{Eq:epsilon_in_new_model}), we obtain
\begin{align}
Y_B = -\frac{cr}{24 \pi v^{2}} 
\left[  \kappa\left(\frac{2A^2}{M_1}\right) - \kappa\left(\frac{A^2}{M_1}\right)  \right] \sum_{J=2}^3  \alpha_{J}^{2}  \sin 2(\theta_{J}-\theta_{1}) \frac{M_1}{M_J} \left( -\frac{3}{2} + \frac{M_1}{M_J} \right),
\end{align}
where we use $F(x) \simeq -3/(2\sqrt{x})$ and $G(x) \simeq -1/x$ for $x \gg 1$.

As benchmark points, we choose $M_1 = 1.5 \times 10^{11}$~GeV, $M_2 = 10 M_1$, $M_3 = 10 M_2$, $A = 1.225$~GeV, $\alpha_2=\alpha_3=5.74A$, $\theta_1=0^\circ$, and $\theta_2=\theta_3=45^\circ$. These values yield $|(M_\nu)_{ij}| = \mathcal{O}(0.01)$~eV from Eq.~(\ref{Eq:MnuSeesaw}), which is consistent with cosmological observations \cite{DESI2025VIJCAP,DESI2025VIIJCAP}, and $|Y_B| = 8.7 \times 10^{-11}$. Hence, for $M_1$ within the two-flavor regime, $10^9~\mathrm{GeV} \lesssim M_1 \lesssim 10^{12}~\mathrm{GeV}$, the derived Dirac mass matrix textures allow for a successful realization of thermal leptogenesis.

\section{Summary \label{section_summary}}
Depending on the mass scale of the lightest right-handed neutrino in the type I seesaw mechanism, leptogenesis proceeds in one of three regimes. Conversely, if we first determine in which of the three regimes leptogenesis occurs, we can determine the mass range of the lightest right-handed neutrino.

In this study we have derived six general textures of the Dirac mass matrix all of which allow leptogenesis to proceed only in the two-flavor regime. If the Dirac mass matrix takes on one of these particular forms, the mass of the lightest right-handed neutrino should be around $10^9 {\rm\: GeV}$ to $10^{12} {\rm \:GeV}$. After deriving the general textures, we presented two specific textures as examples. Our results are expected to contribute to type I seesaw model building efforts where the lightest right-handed neutrino masses are taken to be in the range  $10^9 {\rm\: GeV}$ to $10^{12} {\rm\: GeV}$. We intend to start researching this topic.

Finally we would like to comment that, in addition to the two-flavor regime of vanilla leptogenesis focused on in this work, we are interested in deriving the general textures of the Dirac mass matrix which realize leptogensis in the unflavored and three-flavor regimes as well. We have begun investigations into this subject but have yet to find a simple texture of the mass matrix. Our investigations into this subject matter will continue.

\section*{Acknowledgement}
The authors gratefully acknowledge Michael Fodroci for careful English editing of the manuscript.

\vspace{1cm}

\end{document}